\documentclass[twocolumn,showpacs,preprintnumbers,amsmath,amssymb,pra]{revtex4}
\usepackage{amssymb}

\usepackage{txfonts}
\usepackage{amsmath}

\usepackage{graphicx}
\usepackage{dcolumn}
\usepackage{bm}

\begin{document}


\title{Electromagnetic Field Quantization in Time-Dependent Dielectric Media}

\author{Xiao-Min Bei}
\author{Zhong-Zhu Liu}%
 \email{xiaominbei@gmail.com}
\affiliation{%
Department of Physics, Huazhong University of Science and Technology,
Wuhan, 430074, China
}%

\date{\today}

\begin{abstract}
We present a Gupta-Bleuler quantization scheme for the electromagnetic field in time-dependent dielectric media. Starting from the Maxwell equations, a generalization of the Lorentz gauge condition adapted to time varying dielectrics is derived. Using this gauge, a Gupta-Bleuler approach to quantize all polarizations of the radiation field and the corresponding constraint condition are introduced. This new approach is different from the quantized electromagnetic field in vacuum in the sense that here the contributions of unphysical photons cannot be thoroughly eliminated, which further lead to a surface charge density. Finally, a discussion of potential experimental tests and possible implication is also made.
\end{abstract}

\maketitle

\section{INTRODUCTION}

In quantum field theory, attempts to quantize electromagnetic field in material media date back to more than half a century ago \cite{1,2}. Recently, interests in this problem are mainly related to the flourishing cavity QED \cite{3} and experiments of quantum optics in media \cite{4,5}. In order to solve this problem, various methods have been proposed. Of earlier works, the quantization scheme given by Jauch and Watson in 1948 is taking into account only nondispersive and uniform media \cite{2}. This scheme has been applied to inhomogeneous nondispersive cases too \cite{5}. In recent works, quantization of the electromagnetic field in dispersive media, which can be either homogeneous or inhomogeneous, has been considered \cite{6,7,8}. In particular, there has been attempt to study the most general case with nonlinear materials \cite{9}. More lately, this study has also been extended to anisotropic media \cite{10}. But few of these previous works have taken into account the time-varying properties of the medium.

During the last two decades, it was found that a time-varying dielectric permittivity may lead to generate quanta of electromagnetic field (photons) even from vacuum \cite{11,12,13,14}. This phenomenon is similar to pure quantum effects such as dynamical Casimir effect (DCE) \cite{15}. For this purpose, several experimental schemes have been suggested by the MIR proposal \cite{16}. In one of these schemes, one boundary of a cavity is made of time-dependent media. So, to clarify how to quantize the electromagnetic field in time-dependent media will help to understand these experiments. For this purpose, many literatures focused on one-dimensional case \cite{12}. They split up directly the electromagnetic field into two independent scalar fields assuming to the decoupling of two polarizations \cite{17}. However, this method cannot be directly extrapolated to the quantization of vector fields in three-dimensional cases.

For three-dimensional cases, the four components of the electromagnetic vector potential are not all independent. So we have to fix gauge and to remove those extra degrees of freedom before carrying out the direct quantization of the massless vector field. Generally, the quantization of the electromagnetic field in three-dimensional and time-dependent media is performed in the Coulomb gauge \cite{18,19}.In these theories, two unphysical degrees of freedom of the electromagnetic field are eliminated from the start and only two transverse polarizations are quantized. Recently the path integral quantization of electromagnetic field for dielectrics in covariant gauge has been proposed, where all polarizations of the photon are quantized \cite{20}. Remarkably, it is found that the contributions of the unphysical photons will not be cancelled by those of the ghost fields, which implies that there might be some physical effects of the unphysical photons \cite{21,22}.

In this work, we try to put forward a Gupta-Bleuler quantization procedure to the time-dependent medium. We find that in this case,  the contributions of the unphysical photons cannot be removed completely. Particularly, when the initial state is the vacuum state, contributions from unphysical photons and transverse photons to the average number are equal. This will lead to a detectable surface charge density in the medium surface that was not obtained from theories using Coulomb gauge \cite{18,19} and may be tested experimentally by detecting the existence of this surface charge density. Moreover, if the Gupta-Bleuler theory describes facts more reasonably, it will also provide another possible way to detect the dynamical Casimir effect.

The article is organized as follows.  In Sec. II, we firstly derive a general Lorentz gauge from Maxwell equations in the time-dependent medium. In Sec. III, we start by constructing a new Lagrangian based on this gauge and then present a canonical quantization scheme of the electromagnetic field. The Gupta-Bleuler condition is introduced in Sec. IV. The dielectric permittivity with sinusoidal time dependence is considered in Sec. V, and we show that by the rotating-wave approximation, it could lead to the effect of unphysical photons, namely the surface charge density. Possible implications and experimental tests are discussed in Sec. VI. We then conclude our work in the last section.

Throughout this paper, a signature of (+,-,-,-) for the metric will be adopted. Greek indices are summed over (0,1,2,3) while Latin indices denote the spatial components (1,2,3). However, letter $\lambda$ is singled out as the polarization state label. And natural units $c =\hbar =\varepsilon _0 =\mu _0 = 1$ will be used.

\section{GENERAL GAUGE CONDITION}

In this section we will develop a general Lorentz gauge adaptive to time-varying media. It is convenient to start with the consideration that the pure radiation field and the macroscopic source free Maxwell equations in media take the form
\begin{equation}
\label{eq1}
\nabla\cdot{\textbf{D}} = 0,
\end{equation}
\begin{equation}
\label{eq2}
\nabla\times\textbf{E}+\frac{\partial\textbf{B}}{\partial{t}} = 0,
\end{equation}
\begin{equation}
\label{eq3}
\nabla\cdot{\textbf{B}} = 0,
\end{equation}
\begin{equation}
\label{eq4}
\nabla\times\textbf{H}-\frac{\partial\textbf{D}}{\partial{t}} = 0.
\end{equation}
In the following, we shall restrict our discussions to the case of linear, nondispersive, time-dependent dielectric media, and assume that the constitutive relations between $\left( {\textbf{D},\textbf{B}} \right)$ and $\left( {\textbf{E}},{\textbf{H}} \right)$ can be simply expressed by
\begin{equation}
\label{eq5}
\textbf{D} = \varepsilon (t)\textbf{E},
\end{equation}
\begin{equation}
\label{eq6}
\textbf{B} = \textbf{H},
\end{equation}
where $\varepsilon (t)$ is a time-dependent dielectric permittivity. Strictly speaking, the dielectric permittivity is complex but in this paper we assume that it is real \cite{19}.

Generally the field strength, $\textbf{E}$ and $\textbf{B}$, in quantum mechanics, are regarded as derived quantities and the fundamental 4-vector $A^\mu = \left( {A^0,\textbf{A}} \right)$ can be introduced according to the relations
\begin{equation}
\label{eq7}
\textbf{B} = \nabla \times \textbf{A},
\end{equation}
\begin{equation}
\label{eq8}
\textbf{E} = - \nabla A_0 - \frac{\partial \textbf{A}}{\partial t}.
\end{equation}
With this potential, the two homogeneous Maxwell equations (\ref{eq2}) and (\ref{eq3}) are automatically satisfied, and the other two inhomogeneous equations (\ref{eq1}) and (\ref{eq4}) can be written in terms of Eqs. (\ref{eq7}) and (\ref{eq8}) as
\begin{equation}
\label{eq9}
\mbox{ }\frac{\partial ^2}{\partial t^2}\left( {\varepsilon (t)A_0 } \right)
- \nabla ^2A_0 = \frac{\partial }{\partial t}\left( {\frac{\partial
}{\partial t}\left( {\varepsilon (t)A_0 } \right) +  {\nabla \cdot
\textbf{A}} } \right),
\end{equation}
\begin{equation}
\label{eq10}
\frac{\partial }{\partial t}\left( {\varepsilon (t)\frac{\partial
\textbf{A}}{\partial t}} \right) - \nabla ^2\textbf{A} = - \nabla \left(
{\frac{\partial }{\partial t}\left( {\varepsilon (t)A_0 } \right) + \nabla
\cdot \textbf{A}} \right).
\end{equation}
From these expressions, we can choose a gauge condition by imposing the requirement
\begin{equation}
\label{eq11}
\frac{\partial }{\partial t}\left( {\varepsilon (t)A_0 } \right) + \nabla
\cdot \textbf{A} = 0.
\end{equation}
This gauge condition (\ref{eq11}) is a generalization, appropriate to the presence of time-dependent dielectrics, of the Lorentz gauge condition $\left(\frac{\partial }{\partial t}A_0+\nabla\cdot \textbf{A}=0\right)$. According to this constraint, Eqs. (\ref{eq9}) and (\ref{eq10}) decouple into inhomogeneous damped wave equations, with the time and space components written separately as
\begin{equation}
\label{eq12}
\nabla ^2A_0 - \ddot {\varepsilon }(t)A_0 - 2\dot {\varepsilon}(t)\frac{\partial }{\partial t}A_0 - \varepsilon(t)\frac{\partial^2}{\partial t^2}A_0 = 0,
\end{equation}
\begin{equation}
\label{eq13}
\nabla ^2\textbf{A} - \dot {\varepsilon }(t)\frac{\partial \textbf{A}}{\partial
t} - \varepsilon (t)\frac{\partial ^2}{\partial t^2}\textbf{A} = 0,
\end{equation}
where the dot stands for a time derivate. Generally speaking, we can see the apparent distinction between these two equations, which will lead to different patterns of evolution in time and space. In addition, we can attain negative or positive signs of the term $\dot{\varepsilon }(t)$, which correspond respectively to wave attenuation or amplification. If one makes a special choice $\varepsilon (t) = 1$, both Eq. (\ref{eq12}) and Eq. (\ref{eq13}) will return back to the d'Alembert equation.

\section{CANONICAL QUANTIZATION IN TIME-DEPENDENT MEDIA}

In this section we try to canonically quantize the electromagnetic field $A^\mu \left( x \right)$ in time-dependent media. First of all we observe that Eqs. (\ref{eq12}) and (\ref{eq13}) can be shown as the equations of motion derived from the variational principle with the following Lagrangian
\begin{equation}
\label{eq14}
{{\cal L}}' = - \frac{1}{4}F_{\mu \nu } H^{\mu \nu } - \frac{\zeta
}{2}\left( {\partial _0 \left( {\varepsilon \left( t \right)A^0} \right) +\partial_i A^i} \right)^2,
\end{equation}
where
\begin{equation}
\label{eq15}
H^{\mu \nu } = \left[ {{\begin{array}{*{20}c}
 {\varepsilon ^2\left( t \right)F^{00}} \hfill & {\varepsilon \left( t
\right)F^{0j}} \hfill \\
 {\varepsilon \left( t \right)F^{i0}} \hfill & {F^{ij}} \hfill \\
\end{array} }} \right].
\end{equation}
Here the field strength tensor is defined as $F_{\mu \nu }=\partial_\mu A_\nu - \partial _\nu A_\mu $ and $\zeta $ is a parameter that can be chosen freely. To simply the calculations, we generally set $\zeta = 1$ which is called the Feynman gauge. The extra term on the right hand side of Eq. (\ref{eq14}) is known as gauge-fixing term constructed by the general Lorentz gauge (\ref{eq11}). We find that this modified Lagrangian satisfies canonical quantization prescription. For simplicity we may take$\tilde{A}_\mu=\left(\varepsilon\left(t\right)A_0,A_i\right)$. can be shown as the equations of motion derived from the variational principle with the following Lagrangian (\ref{eq14}) as the follows
\begin{equation}
\label{eq16}
\Pi ^\mu = \frac{\partial {{\cal L}}'}{\partial \left( {\partial _0 \tilde{A}_\mu }\right)} = \left( { - \partial _0 \left({\varepsilon \left( t \right)A^0} \right), - \varepsilon \left( t\right)\partial _0 A^i} \right).
\end{equation}
For now the $\Pi^0$ no longer vanishes and then we can continue to carry out the canonical quantization programme.

As a routine, the quantization of electrodynamics in dielectric media is performed by imposing canonical equal-time commutation relations (ETCR)
\begin{equation}
\label{eq17}
\left[ {\tilde{A}^\mu \left( {\textbf{x},t} \right),\Pi ^\nu \left( {{\textbf{x}}',t}\right)} \right] = ig^{\mu \nu }\delta ^3\left( {\textbf{x} - {\textbf{x}}'}\right),
\end{equation}
\begin{equation}
\label{eq18}
\left[ {\tilde{A}^\mu \left( {\textbf{x},t} \right),\tilde{A}^\nu \left( {{\textbf{ x}}',t}\right)} \right] = \left[ {\Pi ^\mu \left( {\textbf{x},t} \right),\Pi ^\nu\left( {{\textbf{x}}',t} \right)} \right] = 0,
\end{equation}
where $g^{\mu \nu } = g_{\mu \nu } = diag\left( {1, - 1, - 1, - 1} \right)$ is the flat Minkowski metric. Note that all four components of $\tilde{A}^\mu $ and $\Pi ^\mu $ have been brought into consideration and they obey these relations.

In order to characterize quantum properties of the field, the operator
$A_\mu \left( x \right)$ in the Heisenberg picture is expanded in the polarization vectors $e_{k,\mu }^{\left( \lambda \right)} $ as
\begin{equation}
\label{eq19}
{A}_\mu\left(x\right)=\sum_k\frac{1}{\sqrt{2\omega_{kt}}}\sum_{\lambda=0}^3e^{\left(\lambda\right)}_{k,\mu}
\left(a_k^{\left(\lambda\right)}\left(t\right)\phi_{kt}\left(\textbf{x}\right)
+a_k^{\left(\lambda\right)\dag}\left(t\right)\phi_{kt}^*\left(\textbf{x}\right)\right),
\end{equation}
where $a_k^{\left( \lambda \right)} \left( t \right)$ and $a_k^{\left( \lambda\right)\dag } \left( t \right)$ are the annihilation and creation operators. For each $k$, the four polarization 4-vectors $e_{k,\mu }^{\left( 0 \right)}$, $e_{k,\mu }^{\left( 1 \right)} $, $e_{k,\mu }^{\left( 2 \right)} $ and $e_{k,\mu }^{\left( 3 \right)} $ are linearly independent, and are usually chosen to be real. Here $\phi_{kt}\left(\textbf{x}\right)$ and $\omega_{kt}$ are respectively instantaneous mode functions and eigenfrequencies, which obey the equation
\begin{equation}
\label{eq20}
\left(\nabla^2+\varepsilon\left(t\right)\omega_{kt}^2\right)\phi_{kt}\left(\textbf{x}\right)=0.
\end{equation}
The index $t$ is regarded as a parameter indicating that these mode functions and eigenfrequencies depend on the value of $\varepsilon \left(t \right)$ at each time $t$.

In what follow we will confine our consideration to the electromagnetic field inside a three-dimensional cavity filled with time-dependent media. By choosing an orthonormal set of mode functions $\left\{\phi_{kt}\left(\textbf{x}\right)\right\}$, they are required to satisfy the orthonormal conditions \cite{12}
\begin{equation}
\label{eq21}
\int_Vd^3x\ \varepsilon\left(t\right)\phi_{kt}\left(\textbf{x}\right)\phi_{k't}^*\left(\textbf{x}\right)=\delta_{kk'},
\end{equation}
where the integration is performed over volume $V$. In obtaining Eq. (\ref{eq5}) we have assumed that the medium is isotropic so that the mode function may be described by plane wave
\begin{equation}
\label{eq22}
\phi_{kt}\left(\textbf{x}\right)=L^{-3/2}\frac{1}{\sqrt{\varepsilon\left(t\right)}}e^{i\textbf{k}\cdot\textbf{x}},
\end{equation}
with the length of one side of the cavity $L$. Furthermore, the periodic boundary conditions now are read as
\begin{equation}
\label{eq23}
k_1=\frac{2{\pi}n_x}{L},\ k_2=\frac{2{\pi}n_y}{L},\ k_3=\frac{2{\pi}n_z}{L},\ \ n_x,n_y,n_z=0,\pm1,\pm2,\cdot\cdot\cdot.
\end{equation}
Note that in the formal limit $L\to\infty$, these wave vectors become continuous rather than discrete, which describes the electromagnetic field in free space filled with media. Then we can define a 4-vector
\begin{equation}
\label{eq24}
\ell_\mu=\left(\sqrt{\varepsilon\left(t\right)}k_0,k_1,k_2,k_3\right),
\end{equation}
with $k_0=\omega_{kt}$. From (\ref{eq20}), (\ref{eq22}) and (\ref{eq24}) it follows that the energy-momentum relation of photons can be simply written as
\begin{equation}
\label{eq25}
\ell_\mu\ell^\mu=0,
\end{equation}
and eigenfrequencies can be expressed as $\omega_{kt}=\sqrt{k_1^2+k_2^2+k_3^2}/\sqrt{\varepsilon\left(t\right)}$.

At present, to exhibit the obvious physical meanings of the expansion (\ref{eq19}), the polarization vectors $e_{k,\mu }^{\left( \lambda \right)} $ can be chosen in a fixed reference frame as \cite{22}
\begin{equation}
\label{eq26}
e_{k,\mu}^{\left(0\right)}=\left(\begin{array}{c}
                       1 \\
                       0 \\
                       0 \\
                       0
                     \end{array}\right), e_{k,\mu}^{\left(1\right)}=\left(\begin{array}{c}
                                                0 \\
                                              k_2 \\
                                              -k_1 \\
                                              0
                                              \end{array}
                     \right)\frac{1}{k_{\parallel}}, e_{k,\mu}^{\left(2\right)}=\left(\begin{array}{c}
                                                0 \\
                                              k_1k_3 \\
                                              k_2k_3 \\
                                              -k_{\parallel}^2
                                              \end{array}
                     \right)\frac{1}{kk_{\parallel}}, e_{k,\mu}^{\left(3\right)}=\left(\begin{array}{c}
                                                0 \\
                                              k_1 \\
                                              k_2 \\
                                              k_3
                                              \end{array}
                     \right)\frac{1}{k},
\end{equation}
with $k_\parallel=\sqrt{k_1^2+k_2^2}$ and $k=\sqrt{k_\parallel^2+k_3^2}$. These vectors form a set of basis vectors that satisfy the following orthonormal and complete relations
\begin{equation}
\label{eq27}
e_{k,\mu}^{\left(\lambda\right)}{\cdot}e_{k,\mu}^{\left(\lambda'\right)}=g^{\lambda\lambda'},\ \ \ \ \sum_{\lambda\lambda'}g^{\lambda\lambda'}e_{k,\mu}^{\left(\lambda\right)}e_{k,\nu}^{\left(\lambda'\right)}=g_{\mu\nu},
\end{equation}
where $g^{\lambda {\lambda }'} = diag\left( {1, - 1, - 1, - 1} \right)$, like $g^{\mu \nu }$. It is seen from Eqs. (\ref{eq24}) and (\ref{eq26}) that the polarization vectors with $\lambda = 1,2$ are transversal and satisfy $\ell \cdot e_k^{\left( {\lambda = 1,2} \right)} = 0$ while those with $\lambda =0,3$ are timelike and longitudinal constrained by $\ell \cdot e_k^{\left( 0
\right)} = - \ell \cdot e_k^{\left( 3 \right)}$.

Substituting Eqs. (\ref{eq26}) and (\ref{eq22}) into the expansion (\ref{eq19}) we obtain the standard form
\begin{eqnarray}
\label{eq28}
A_\mu\left(x\right)=\frac{1}{L^{3/2}}\sum_k\frac{e^{-ik_jx^j}}{\sqrt{2k}}\times
\left\{e_{k,\mu}^{\left(0\right)}a_k^{\left(0\right)}\left(t\right)\varepsilon^{-\frac{3}{4}}\left(t\right)\right.\nonumber\\
\left.+e_{k,\mu}^{\left(3\right)}a_k^{\left(3\right)}\left(t\right)\varepsilon^{-\frac{1}{4}}\left(t\right)
+e_{k,\mu}^{\left(1\right)}a_k^{\left(1\right)}\left(t\right)\varepsilon^{-\frac{1}{4}}\left(t\right)\right.\nonumber\\
\left.+
e_{k,\mu}^{\left(2\right)}a_k^{\left(2\right)}\left(t\right)\varepsilon^{-\frac{1}{4}}\left(t\right)\right\}+c.c.,
\end{eqnarray}
with $j=1,2,3$. Here the polarizations $\lambda = 1,2$ correspond to the transversal parts of the potential (\ref{eq21}) which can be also given in Coulomb gauge. We then insert Eq.(\ref{eq28}) into the wave equations (\ref{eq12}) and (\ref{eq13}), and obtain the evolution equations of $a_k^{\left( \lambda \right)}\left( t \right)$ and $a_k^{\left( \lambda \right) \dag } \left( t \right)$
\begin{subequations}\label{eq:29}
\begin{align}
\frac{da_k^{\left(\lambda\right)}\left( t \right)e^{-ik_jx^j}}{dt}=-i\omega_{kt}a_k^{\left(\lambda\right)}\left( t\right)e^{-ik_jx^j}\pm\frac{1}{2}\frac{\dot{\omega_{kt}}}{\omega_{kt}}a_k^{\left( \lambda\right)\dag}\left( t \right)e^{ik_jx^j},\label{eq:29a}\\
\frac{da_k^{\left(\lambda\right)\dag}\left( t \right)e^{ik_jx^j}}{dt}=i\omega_{kt}a_k^{\left(\lambda\right)\dag}\left(t\right)
e^{ik_jx^j}\pm\frac{1}{2}\frac{\dot{\omega_{kt}}}{\omega_{kt}}a_k^{\left( \lambda\right)}\left( t \right)e^{-ik_jx^j},\label{eq:29b}
\end{align}
\end{subequations}
where the upper sign (+) in the right hand side of the equations denotes timelike photon $\lambda=0$, and the lower sign (-) denotes transverse and longitudinal photons $\lambda=1,2,3$. Now, at the initial time $t=0$, we assume $a_k^{(\lambda)}\left(0\right)=a_k^{(\lambda)}$ to be usual annihilation operators. If we set the parameter $\varepsilon\left(t\right)=1$, then all coefficients $\dot{\omega}_{kt}$ are zero. In this special case, Eq. (\ref{eq:29a}) reduces to $\frac{da_k^{(\lambda)}\left(t\right)}{dt}=-i\omega_{kt}a_k^{(\lambda)}\left(t\right)$. Hence the time evolution of the annihilation operators is simply expressed by $a_k^{(\lambda)}\left(t\right)=a_k^{(\lambda)}e^{-i\omega_kt}$ with a time-independent frequency $\omega_k$. Similarly, we can get the time evolution of the creation operators in the same way. That is to say, they have been reduced to the expression in vacuum \cite{23}.

Now we put Eq. (\ref{eq28}) into the commutation rules (\ref{eq17}) and (\ref{eq18}), and then derive the commutation relations for the components $a_k^{\left( \lambda\right)} \left( t \right)$ and $a_k^{\left( \lambda \right) \dag} \left( t\right)$ as
\begin{equation}
\label{eq30}
\left[a_k^{\left(\lambda\right)}\left(t\right),a_{k'}^{\left(\lambda'\right)\dag}\left(t\right)\right]
=-g^{\lambda\lambda'}\delta^3\left(\textbf{k}-\textbf{k}'\right),
\end{equation}
\begin{equation}
\label{eq31}
\left[a_k^{\left(\lambda\right)}\left(t\right),a_{k'}^{\left(\lambda'\right)}\left(t\right)\right]
=\left[a_k^{\left(\lambda\right)\dag}\left(t\right),a_{k'}^{\left(\lambda'\right)\dag}\left(t\right)\right]=0.
\end{equation}
It can be seen that for longitudinal and transverse photons $\lambda=1,2,3$ these relations are subject to the ordinary bosonic quantization scheme. Nevertheless, for timelike photons $\lambda=0$ there is an extra minus sign on the right-hand side, which would bring about a remarkable consequence that the Hilbert space of the photon field carries an indefinite metric.

Then, for the operators $a_k^{\left( \lambda \right)} \left( t \right)$ and $a_k^{\left( \lambda \right) \dag } \left( t \right)$, Heisenberg equations of motion are given by
\begin{eqnarray}
\label{eq32}
i\dot{a}_k^{\left( \lambda \right)}\left( t \right)= \left[a_k^{\left( \lambda
\right)} \left( t \right),H_{eff} \right], \nonumber\\
i\dot{a}_k^{\left( \lambda \right) \dag }\left( t \right) = \left[
a_k^{\left( \lambda \right)\dag}\left( t \right),{H}_{eff} \right].
\end{eqnarray}
Comparing the above equations (\ref{eq32}) with the evolution equations (\ref{eq:29}), the effective Hamiltonian operator in the Heisenberg picture is found to be
\begin{eqnarray}
\label{eq33}
{H}_{eff}=\sum_k\sum_{\lambda=0}^3\left[\left(-g_{\lambda\lambda}\right)
\omega_{kt}a_k^{\left(\lambda\right)\dag}\left(t\right)a_k^{\left(\lambda\right)}\left(t\right)\right.\nonumber\\
\left.+\frac{i}{4}\frac{\dot{\kappa}\left(t\right)}{\kappa\left(t\right)}
\left(a_k^{\left(\lambda\right)\dag}\left(t\right)a_k^{\left(\lambda\right)\dag}\left(t\right)
-a_k^{\left(\lambda\right)}\left(t\right)a_k^{\left(\lambda\right)}\left(t\right)\right)\right],
\end{eqnarray}
\noindent
with $\kappa \left( t \right) = \sqrt {\varepsilon \left( t \right)} $. This Hamiltonian in the Schr\"{o}dinger picture is given simply as \cite{24}
\begin{eqnarray}
\label{eq34}
{H}_{eff}^S=\sum_k\sum_{\lambda=0}^3\left[\left(-g_{\lambda\lambda}\right)
\omega_{kt}a_k^{\left(\lambda\right)\dag}a_k^{\left(\lambda\right)}\right.\nonumber\\
\left.+\frac{i}{4}\frac{\dot{\kappa}\left(t\right)}{\kappa\left(t\right)}
\left(a_k^{\left(\lambda\right)\dag}a_k^{\left(\lambda\right)\dag}
-a_k^{\left(\lambda\right)}a_k^{\left(\lambda\right)}\right)\right],
\end{eqnarray}
We note that in the formal limit $\varepsilon\left(t\right)\rightarrow1$ the Hamiltonian (\ref{eq34}) turns into the Hamiltonian for the electromagnetic field in vacuum \cite{23}. For performing the perturbation theory we would like to further investigate the system's evolution in the interaction picture. Through the unitary transformation $U_0\left(t\right)=\exp{\left(-i\int_0^t d\tau H_0\left(\tau\right)\right)}$, we get the system Hamiltonian in the interaction picture as \cite{25}
\begin{eqnarray}
\label{eq34'}
{H}^I=H_0^I+H_1^I,
\end{eqnarray}
where the free part is $H_0^I=\sum_k\sum_{\lambda=0}^3\left(-g_{\lambda\lambda}\right)\omega_{kt}a_k^{\left(\lambda\right)\dag}a_k^{\left(\lambda\right)}$, and interaction Hamiltonian is
\begin{eqnarray}
\label{eq35}
{H}_1^I=\sum_k\sum_{\lambda=0}^3\frac{i}{4}\frac{\dot{\kappa}\left(t\right)}{\kappa\left(t\right)}
\left(a_k^{\left(\lambda\right)\dag}a_k^{\left(\lambda\right)\dag}e^{i2\Omega_k\left(t\right)}
-a_k^{\left(\lambda\right)}a_k^{\left(\lambda\right)}e^{-i2\Omega_k\left(t\right)}\right),
\end{eqnarray}
with $\Omega_k\left(t\right)=\int_0^t\omega_{k\tau}d\tau$. It is seen that the first term in Eq.(\ref{eq35}) is two-photon process characterized by the term $a_k^{\left(\lambda\right)\dag}a_k^{\left(\lambda\right)\dag}$, so that photon pairs can be created from the vacuum state. Secondly, in contrast to \cite{25,26}, we are interested here in whether it is likely to create longitudinal and timelike photons, in addition to transverse photons. This will be discussed in detail in Sec.V.

\section{THE GUPTA-BLEULER METHOD IN TIME-DEPENDENT MEDIA}

So far we have not been really dealing with Maxwell's theory in media, since we have adjusted the Lagrangian (\ref{eq14}). To recover it, let us now turn our attention to the general Lorentz gauge (\ref{eq11}). However, after the quantization we cannot simply take this gauge condition as an operator identity. Comparing with the classical Gupta-Bleuler method \cite{27,28}, we impose the general gauge condition on physical states $|\phi\rangle$ as the expectation value
\begin{equation}
\label{eq36}
\langle\phi|\left[\partial_0\left(\varepsilon\left(t\right)A^0\right)+\partial_jA^j\right]|\phi\rangle=0.
\end{equation}
For simplicity, the above formulas can be equivalently expressed as
\begin{equation}
\label{eq37}
\left(\partial_0\left(\varepsilon\left(t\right)A^0\right)+\partial_jA^j\right)^{\left(+\right)}|\phi\rangle=0,
\end{equation}
where only the destruction part of this gauge condition is allowed to act on physical states. Now, we insert (\ref{eq28}) into (\ref{eq37}) and take into account of $\ell \cdot e_k^{\left( {\lambda = 1,2} \right)} = 0$ and $\ell \cdot e_k^{\left( 0\right)} = - \ell \cdot e_k^{\left( 3 \right)}$.

Then, the constraint at the initial time can be rewritten as
\begin{equation}
\label{eq38}
\left[-a_k^{\left( 0 \right)}+ a_k^{\left( 3 \right)}\right]|\phi\rangle_0=0,
\end{equation}
where $|\phi\rangle_0$ is the initial quantum state (at $t=0$). This expression provides a link between the timelike and longitudinal photons. And from Eq. (\ref{eq38}) and its adjoint we have
\begin{equation}
\label{eq39}
_0\langle\phi|a_k^{\left(3\right)\dag}a_k^{\left(3\right)}|\phi\rangle_0-\
_0\langle\phi|a_k^{\left(0\right)\dag}a_k^{\left(0\right)}|\phi\rangle_0=0.
\end{equation}
By this means we will remove the contributions of the unphysical photons to the expectation value of the number operator and energy in the initial time. However, with the time evolution, we will find that in the medium the contributions of the longitudinal and timelike photons cannot be cancelled out by each other. In this case, it is observed  that there is a surface charge density in the medium surface. We will devote a full discussion to this subject in  the next section.

\section{SURFACE CHARGE DENSITY}

In the following, by means of dielectric permittivity changing in periodic time dependence, we will study the nature of electromagnetic fields in the medium and investigate the possible impact on the detection of the dynamical Casimir effect .
Now we assume sinusoidal time-varying permittivity
\begin{equation}
\label{eq40}
\varepsilon\left(t\right)=\varepsilon+2\delta\sin{\left(2{\Omega}t\right)},
\end{equation}
where $\varepsilon$ is a constant, the amplitude $\delta/\varepsilon\ll1$, and $\Omega=\omega_1$ is the static value of the fundamental mode for $\varepsilon\left(t\right)=\varepsilon$ \cite{12}. Then  by taking dielectric permittivity (\ref{eq40})into the Hamiltonian operator $H_1^I$ (\ref{eq35}) and using the rotating-wave approximation \cite{25,26}, we can obtain
\begin{eqnarray}
\label{eq41}
H_1^I\approx\frac{i}{4}\frac{\delta}{\varepsilon}\Omega\sum_{\lambda=0}^3
\left(a_1^{\left(\lambda\right)\dag}a_1^{\left(\lambda\right)\dag}-
a_1^{\left(\lambda\right)}a_1^{\left(\lambda\right)}\right).
\end{eqnarray}
Generally assuming that the electromagnetic field is initially in thermal equilibrium states at finite temperature $T$ \cite{26}, the expectation value of the number operator in the $k$th mode can be written in the form
\begin{eqnarray}
\label{eq42}
\langle N_k\left(t\right)\rangle&&\approx\left(-g_{\lambda\lambda}\right){\langle}\sum_{\lambda=0}^3\exp{\left(iH_1^It\right)}
a_k^{\left(\lambda\right)\dag}a_k^{\left(\lambda\right)}\exp{\left(-iH_1^It\right)}{\rangle}_0\nonumber\\
&&=\left(-g_{\lambda\lambda}\right)\sum_{\lambda=0}^3\left({\langle}a_k^{\left(\lambda\right)\dag}a_k^{\left(\lambda\right)}{\rangle}_0\right.\nonumber\\
&&\ \ \ \ \ \ \ \ \ \ \ \left.+\delta_{k1}\sinh^2\left(\frac{\delta}{2\varepsilon}{\Omega}t\right)
{\langle}a_k^{\left(\lambda\right)\dag}a_k^{\left(\lambda\right)}+
a_k^{\left(\lambda\right)}a_k^{\left(\lambda\right)\dag}{\rangle}_0\right)\nonumber\\
&&=\sum_{\lambda=0}^3{\langle}N_k^{\left(\lambda\right)}{\rangle}_0
+\delta_{k1}\sum_{\lambda=0}^3
\left[\sinh^2\left(\frac{\delta}{2\varepsilon}{\Omega}t\right)
\left(1+2{\langle}N_1^{\left(\lambda\right)}{\rangle}_0\right)\right].\nonumber\\
\end{eqnarray}
Because of the formula (\ref{eq39}), $\langle{N_k^{(3)}}\rangle_0+\langle{N_k^{(0)}}\rangle_0=0$, we get
\begin{eqnarray}
\label{eq42'}
\langle N_k\left(t\right)\rangle\approx\sum_{\lambda=1}^2\langle N_k^{\left(\lambda\right)}\rangle_0
&&+\delta_{k1}\left[\sum_{\lambda=1}^2\left(\sinh^2\left(\frac{\delta}{2\varepsilon}\Omega t\right)\left(1+2\langle N_1^{\left(\lambda\right)}\rangle_0\right)\right)\right.\nonumber\\
&&\left.+\sum_{\lambda=0,3}\sinh^2\left(\frac{\delta}{2\varepsilon}\Omega t\right)\right].
\end{eqnarray}
The first term $\sum_{\lambda=1}^2\langle N_k^{\left(\lambda\right)}\rangle_0$ denotes the initial photons contain only the transverse photons, while the second term represents the average number of photons created grows exponentially with time. It is noticeable that here the contributions of the longitudinal and timelike photons are nonzero. Nevertheless, only the transverse part of photon generation $\lambda=1,2$ has been enhanced by the initial thermal photons. This means that the number of created unphysical photons $\lambda=0,3$ is much smaller than that of the transverse photons at nonzero temperature.

As a next step, we let the vacuum state $|0\rangle$ be the initial quantum state (at $T=0$), then the total number of photons created from vacuum in the time $t$
\begin{eqnarray}
\label{eq43}
\langle0|N\left(t\right)|0\rangle=\sum_{\lambda=0}^3\sinh^2\left(\frac{\delta}{2\varepsilon}\Omega t\right).
\end{eqnarray}
According to this result, we find that all polarization components have the same contribution to the number of photons generated from vacuum. This is to say, the number of unphysical photons is equal to those of transverse photons in this case.

Further, we would exhibit the content of the electric and magnetic fields. Through $E_i\left(x\right)=\frac{\partial}{\partial x^0}A_i-\frac{\partial}{\partial x^i}A_0$ we can obtain from Eq.(\ref{eq28})
\begin{eqnarray}
\label{eq44}
E_i\left(x\right)=&&-\frac{1}{L^{3/2}}\sum_k\frac{e^{-ik_jx^j}\varepsilon^{-3/4}\left(t\right)}{\sqrt{2k}}\nonumber\\
&&\times\left\{
\left(
\begin{array}{c}
k_1 \\
k_2 \\
k_3 \\
\end{array}
\right)
i\left(-a_k^{\left(0\right)}\left(t\right)+a_k^{\left(3\right)}\left(t\right)\right)\right.\nonumber\\
&&\left.+\left(\begin{array}{c}
          k_2 \\
          -k_1 \\
          0
        \end{array}\right)\frac{ik}{k_{||}}a_k^{\left(1\right)}\left(t\right)
+\left(\begin{array}{c}
         k_1k_3 \\
         k_2k_3 \\
         -k_{||}^2
       \end{array}
\right)\frac{i}{k_{||}}a_k^{\left(2\right)}\left(t\right)
\right\}+c.c.,\nonumber\\
\end{eqnarray}
and from $B_i\left(x\right)=\varepsilon_{ijk}\partial_jA_k$ for the magnetic field
\begin{eqnarray}
\label{eq45}
B_i\left(x\right)=&&\frac{1}{L^{3/2}}\sum_k\frac{e^{-ik_jx^j}\varepsilon^{-1/4}\left(t\right)}{\sqrt{2k}}\nonumber\\
&&\times\left\{
\left(
\begin{array}{c}
k_1k_3 \\
k_2k_3 \\
-k_{||}^2 \\
\end{array}
\right)\frac{-i}{k_{||}}a_k^{\left(1\right)}\left(t\right)
+\left(\begin{array}{c}
          k_2 \\
          -k_1 \\
          0
        \end{array}\right)\frac{ik}{k_{||}}a_k^{\left(2\right)}\left(t\right)
\right\}+c.c..\nonumber\\
\end{eqnarray}
It is seen that longitudinal and timelike photons exist only in the longitudinal component of the electric field, not in the transverse electric and the magnetic fields.

Therefore we only need to investigate the effects of the longitudinal electric field. In contrast to \cite{29}, the energy of the unphysical photons can be represented from (\ref{eq44}) as
\begin{eqnarray}
\label{eq46}
\frac{1}{2}\int d^3x\varepsilon\left(t\right)E_{//}^2=\sum_k\omega_{kt}\left[a_k^{\left(3\right)\dag}\left(t\right)a_k^{\left(3\right)}\left(t\right)
-a_k^{\left(0\right)\dag}\left(t\right)a_k^{\left(0\right)}\left(t\right)\right],\nonumber\\
\end{eqnarray}
where $E_{//}$ is the longitudinal part of the electric field operator $E_i\left(x\right)$. At the same time, from \cite{30} this longitudinal electric field on either side of the boundary surface satisfies the boundary condition
\begin{eqnarray}
\label{eq47}
\left(\varepsilon\left(t\right)\textbf{E}_{//}^\varepsilon\right)-\textbf{E}_{//}^{vacuum}=\sigma,
\end{eqnarray}
where $\sigma$ is the surface charge density. In vacuum, the longitudinal component of the electric field $\textbf{E}_{//}^{vacuum}$ is null \cite{29}. Finally, using Eq. (\ref{eq43}) and the expectation value of (\ref{eq46}), we get
\begin{eqnarray}
\label{eq48}
\sigma\approx2\sinh\left(\frac{\delta}{2\varepsilon}\Omega t\right)\frac{\varepsilon^{1/4}\sqrt{\pi}}{L^2}.
\end{eqnarray}
Note that when the dielectric permittivity varies with time, the contributions of the longitudinal and timelike photons will no longer vanish, which lead to a detectable surface charge density.

This result has not been obtained from the previous quantum theory \cite{18,19}. Thus, in time-dependent media, the Gupta-Bleuler quantization scheme and the quantization in Coulomb gauge are no equivalent. Theoretically it is impossible to tell which quantum theory is more reasonable, but experimentally this may be tested through the measurement of  the surface charge density. If the Gupta-Bleuler theory is reasonably in this case, we can also propose another experimental method for the detection of the dynamical Casimir effect.

In the experiment \cite{15,16}, one may accomplish the wall motion by manipulating the dielectric permittivity of the medium in the cavity. And from above mentioned reviews the unphysical photon created from time dependent media may lead to a surface charge density. Let us insert some explicit number, and estimate this surface charge density. For instance, we use the same experimental parameters considered in Ref.\cite{16}: $\delta/\varepsilon\sim10^{-8}$ and the frequency of the lowest mode is $\omega_1\sim GHz$ for $L_0\sim0.1m$, then approximately $\sigma\approx10^{-14}C\cdot m^{-2}$ will be created in the cavity during $1s$.

Though detecting this surface charge density, we can verify whether the unphysical photons make a nonzero contribution to observable quantities in time-dependent media. Namely, are the quantization in the Lorentz gauge and that in the Coulomb gauge completely equivalent in this case?

At the same time, from the DCE \cite{16} detection experiment, it is difficult to distinguish between the background of thermal photons and the creation of transverse photons. Therefore, we suggest to detect the surface charge density, hopefully it will provide another probability to observe the dynamical Casimir effect in the near future.

\section{CONCLUSIONS}

Starting from the source free Maxwell equations in media, we presented a gauge condition, which is a generalization of the Lorentz gauge condition appropriate to time varying media. Two different wave equations were obtained in the directions of time and space. It is important to notice that $\dot{\varepsilon}(t)$ is in general nonzero, which may lead to photon generation even from vacuum.

Through the general gauge (\ref{eq11}) a modified Lagrangian was given in Sec. III, which makes $\Pi^0$ no longer vanish, and carries out the canonical quantization procedure. Assuming that the mode functions can be described by plane wave, and using the polarization vectors $e_{k,\mu}^{\left( \lambda \right)}$, the standard expansion of the operator ${A}_\mu \left( x \right)$ and the evolution equations of $a_k^{\left( \lambda \right)} \left( t \right)$ and $a_k^{\left( \lambda\right) \dag } \left( t \right)$ were deduced. Based on the Heisenberg equations of motion, the effective Hamiltonian was immediately derived. But this Hamiltonian formula (\ref{eq33}) suggests that all polarization components contribute to the photon creation. In accordance with the Gupta-Bleuler method in time-dependent media, imposing the constraint condition (\ref{eq11}) on the physical state, we were able to get rid of contributions from the longitudinal and the timelike photons at the initial time.

Finally, by assuming sinusoidal time varying permittivity, we were able to simulate this effect. It is shown that not only the transverse photons but also the longitudinal and timelike photons were obtained in the experiments of DCE. Particularly, for an initial vacuum state, all polarization components have the same contribution to the number of photon generation. Therefore, we can get a detectable surface charge density in the medium surface, and provide another experimental trick for confirming one of the fundamental predictions of the DCE.

\begin{acknowledgments}
This work is supported in part by the National Natural Science Foundation of China (Grant No. 2010CB832800 ).
\end{acknowledgments}


\end{document}